\documentclass[a4paper]{amsart}
\usepackage[utf8]{inputenc} % For UTF-8 encoding
\pdfoutput=1
\usepackage{amsfonts,amsmath,amssymb,amsthm}
\usepackage{mathtools}

\usepackage{graphicx}
\usepackage{tikz-cd}
\usepackage{xcolor}
\usepackage{mdframed}

\usepackage{enumerate}
\usepackage{soul}
\usepackage{listings}
\usepackage[color=red!40]{todonotes}

\usepackage[
    left=1.5in,
    right=1.5in,
    bottom=1.2in
]{geometry}

\definecolor{light-gray}{gray}{0.95} % The shade of grey that Stack Exchange uses
\definecolor{cadmiumgreen}{rgb}{0.0, 0.42, 0.24}

\usepackage[
    backrefs,
    msc-links,
    nobysame,
    lite,
]{amsrefs}

\usepackage{hyperref}
\hypersetup{
    colorlinks,
    citecolor=cadmiumgreen,
    pagebackref,
    pdfauthor={Rohan Pandey, Kai Poffenbarger},
    pdftitle={Glioma Oncostream Birth-Death \& Cytochalasin D Effects},
}

\title[Glioma Oncostream Birth-Death \& Cytochalasin D Effects]{Analyzing the birth-death model of Oncostreams in Glioma, and the effects of Cytochalasin D treatment}
\author{Rohan Pandey}
\author{Kai Poffenbarger}
\date{\today}

\begin{document}

\maketitle

\begin{abstract}
This research project investigates the critical role of oncostreams in glioma aggressiveness, leveraging advanced ex-vivo 3D explants and in-vivo intravital imaging techniques to establish a direct correlation between oncostream density and cancer severity. The primary objective is to model the cell populations within oncostreams, with a specific focus on GFP+ NPA cells, to simulate cancer dynamics and provide insights into tumor behavior. The study employs a simple Birth-Death process to analyze cell population dynamics and treatment effects, building and solving Kolmogorov equations to predict changes in cell population over time.

While the model could be expanded to include additional modulators such as morphological attributes and neurotransmitter exposure, the focus remains on cell population to maintain feasibility. The study also examines various treatment methods, finding that glutamate increases glioma cell movement while histamine reduces it. Collagenase treatment effectively dismantles oncostreams, suggesting a potential therapeutic strategy. For this paper, we specifically are going to be looking at Cytochalasin D, which shows promise in disrupting oncostreams and reducing glioma invasiveness. By integrating these treatment variables into the model, the research aims to understand their impact on glioma cell density within the oncostreams and aggressiveness, thereby contributing to improved cancer management strategies. This comprehensive approach is expected to enhance our understanding of glioma progression and inform the development of effective therapeutic interventions.
\end{abstract}

\section*{Acknowledgements}

We would like to acknowledge and thank Professor Konstantinos Mamis for guiding through the course of this project.

\newpage

\tableofcontents

\newpage
\section{Introduction}

Gliomas are among the most aggressive and deadly forms of brain tumors, posing significant challenges in both treatment and prognosis due to their highly invasive nature and resistance to conventional therapies. These tumors infiltrate the brain parenchyma, making complete surgical removal nearly impossible and contributing to high recurrence rates. Understanding the underlying mechanisms that drive glioma progression and invasiveness is crucial for developing more effective therapeutic strategies. One emerging area of interest is the role of oncostreams, specialized cellular structures within the tumor microenvironment that have been implicated in promoting glioma aggressiveness. This research project aims to investigate the critical role of oncostreams in glioma aggressiveness, leveraging advanced ex-vivo 3D explants and in-vivo intravital imaging techniques to establish a direct correlation between oncostream density and cancer severity.

The primary objective of this study is to model the cell population within oncostreams, focusing specifically on GFP+ NPA cells, which are known to play a significant role in tumor dynamics. By simulating cancer dynamics through this modeling, the study seeks to provide deeper insights into the behavior and progression of gliomas. The methodology employed involves a simple Birth-Death process to analyze cell population dynamics and treatment effects. This approach includes building and solving Kolmogorov equations to predict changes in cell population over time, offering a quantitative framework to understand how oncostream density impacts glioma progression.

While the current model focuses on cell population response to treatment, to maintain feasibility, it is designed with the potential for future expansion to include additional modulators such as morphological attributes and neurotransmitter exposure. This would allow for a more comprehensive understanding of the factors influencing glioma behavior. The study also explores various treatment methods to assess their impact on glioma cell movement and oncostream integrity. For instance, it has been observed that glutamate increases glioma cell movement, potentially exacerbating tumor spread, while histamine has the opposite effect, reducing cell movement and possibly hindering tumor progression. Additionally, collagenase treatment has been found to effectively dismantle oncostreams, suggesting a potential therapeutic strategy for disrupting these critical structures within the tumor microenvironment.

Moreover, the study identifies Cytochalasin D as a promising agent for disrupting oncostreams and reducing glioma invasiveness. By integrating these treatment variables into the model, the research aims to understand their impact on glioma cell population and aggressiveness comprehensively. This integration is expected to yield valuable insights into the mechanisms of glioma progression and the potential for targeted therapeutic interventions. Ultimately, this research seeks to contribute to improved cancer management strategies by enhancing our understanding of glioma dynamics and identifying effective treatment approaches.

In conclusion, this comprehensive approach to studying oncostreams in gliomas, using advanced imaging techniques and mathematical modeling, represents a significant step forward in our quest to unravel the complexities of glioma progression. By focusing on the critical role of oncostreams and the impact of various treatments, this study aims to inform the development of more effective therapeutic interventions, potentially improving outcomes for patients with this devastating disease.

\newpage

\section{Introduction of the Mathematical Model}

\subsection{Simple Birth-Death Model}

In a birth-death processes the appearance of new individual bacterial cells, as well as the disappearance of existing individual bacterial cells play a very important role. Looking into simple birth and simple death processes will allow us to build a basis to expand for further applications going forward. We introduce important values and notation as follows below.

\begin{align*}
    \lambda &\rightarrow \text{birth rate}\\
    \mu &\rightarrow \text{death rate}\\
    N(t) \text{ and } n &\rightarrow \text{population}\\
    \lambda N(t)\Delta t &\rightarrow \text{chance of +1 (1 birth)}\\
    \mu N(t) \Delta t &\rightarrow \text{chance of -1 (1 death)}\\
    1-(\lambda + \mu)N(t)\Delta t &\rightarrow \text{chance of 0 (no change)}\\
    a &\rightarrow \text{Initial Population of Glioma Cells}\\
\end{align*}

Given a birth rate of $\lambda$ of a population of size $N(t)$ with $n$ individuals, we can now create a Kolmogorov equation, which can be solved and used to model this phenomena.

\begin{align}
    p_n(t+\Delta t) &= p_{n-1}(t)\lambda(n-1)\Delta t + \mu(n+1)p_{n+1}(t)\Delta t - \left(\lambda+\mu\right)n\Delta tp_n(t)
\end{align}

Gives us the difference equation:

\begin{align}
    \frac{dp_n(t)}{dt}&=\lambda(n-1)p_{n-1}(t) +\mu(n+1)p_{n+1}(t)-(\lambda+\mu)np_n(t)
\end{align}

Notice that if the population becomes $0$ then no births can occur, so we are only concerned about the process starting with a non-zero population suppose it's $N(0) = a$, leading to an initial condition: \[
p_a(0) = 1
\]

Some things to note here:

\begin{align*}
        \mu &= 0:\\
        \frac{dp_n(t)}{dt} &= \lambda (n-1) p_{n-1}(t) - \lambda n p_n(t)\\
        \lambda &= 0:\\
        \frac{dp_n(t)}{dt} &= \mu (n+1)p_{n+1}(t) - \mu n p_n(t)
\end{align*}

Now using steps shown in \ref{appendix:derivation}

we get the probability generating function for $\lambda = \mu$:

\begin{align}
    G(s,t)&=\left(\frac{1-(\lambda t-1)(s-1)}{1-\lambda t(s-1)}\right)^a
\end{align}
\newpage
and for the probability generating function for $\lambda \ne \mu$

\begin{align}
    G(s,t)=\left(\frac{\mu \omega(s,t)-1}{\lambda \omega(s,t)-1}\right)^a \text{ , where }\omega(s,t) = \frac{(s-1)e^{(\lambda-\mu)t}}{\lambda s - \mu}
\end{align}\\

And we can then represent the value of $p_n(t)$ as a specific distribution:

\begin{align*}
    p_n(t)&=\frac{(\lambda t)^{n-1}}{(1+\lambda t)^{n+1}}\text{, }n\ge1\\
    p_0(t)&=\frac{\lambda t}{1 + \lambda t}
\end{align*}\\

We can follow a similar derivation process for $\lambda\ne\mu$ for $a=1$, which results in:

\begin{align*}
    p_n(t)&=(1-\alpha)(1-\beta)\beta^{n-1}\text{, \space }\alpha=\frac{\mu(e^{(\lambda-\mu)t}-1)}{\lambda e^{(\lambda-\mu)t}-\mu}\text{, \space }\beta=\frac{\lambda(e^{(\lambda-\mu)t}-1)}{\lambda e^{(\lambda-\mu)t}-\mu}\text{, \space }n\ge1\\
    p_0(t)&=\alpha
\end{align*}

\subsection{Cytochalasin D Treatment} \label{Cytochalasin D Treatment}
We now introduce a Population model of the Cytochalasin D treatment and also a model for the concentration of the Cytochalasin D used to help dismantle the oncostreams to slow the glioma invasiveness. And analyze how and whether the growth of glioma changes.\\

\textbf{Drug Background information:} Cytochalasin D is a drug that is known to impede actin polymerization, and was observed to disrupt the structural organization of oncostreams in GFP+ NPA glioma cells. According to \cite{HUANG} the half life of natural Cytochalasin D is 10 minutes and liposomal Cytochalasin D is 4 hours (240 minutes). Because the paper doesn't mention if it uses natural or liposomal Cytochalasin D, we will analyze our model using both types.

\begin{align*}
    \gamma &\rightarrow \text{Decay rate of Cytochalasin D}\\
    \beta &\rightarrow \text{Constant used to scale the effect of the treatment concentration}\\
    C(t) &\rightarrow \text{Concentration of the Cytochalasin D drug administered}\\
    (\mu + C(t))N(t) \Delta t &\rightarrow \text{$-1$ cell implying the death of a cell}\\
    1 - (\lambda + \mu + C(t))N(t) \Delta t &\rightarrow \text{No change in the net population or growth of the cells}
\end{align*}

For the simplicity of the derivation process, we can assign $\beta = 1$. But later, we will change it to be $\beta=\frac{1}{100}$, for reasons explained in \ref{parameter values}. This will not cause any problems with our derivation because $\beta$ is a constant.\\

We can now introduce elementary models to model the Population and Concentration of the Cytochalasin D drug. We are assuming that the treatment is being administered at a constant rate in our paper:

\begin{align}
    \frac{dC(t)}{dt} &= -\gamma C(t)\\
    \frac{dN(t)}{dt} &= \lambda N(t) - \mu N(t) - C(t)N(t)
\end{align}

\vspace{2em}

We now solve the ODEs to get a more accurate form of the $N(t)$ and $C(t)$ equations:

\begin{align*}
    \frac{dC(t)}{dt} &= -\gamma C(t)\\
    \int \frac{dC(t)}{C(t)} &= \int -\gamma dt\\
    \ln{C(t)} &= -\gamma t + A\\
    \text{Using the initial condition of } C(0)&=C_0:\\
    C(t) &= C_0 e^{-\gamma t}
\end{align*}

Using the this result we can solve for the population function. We can use the assumption that $N(0) = a$, where $a$ is the initial population of glioma cells.

\begin{align*}
    \frac{dN(t)}{dt} &= \lambda N(t) - \mu N(t) - C_0e^{-\gamma t}N(t)\\
    \frac{dN(t)}{dt} &= N(t) \left[ \lambda - \mu - C_0e^{-\gamma t} \right]\\
    \int \frac{dN(t)}{N(t)} &= \int [\lambda - \mu - C_0e^{-\gamma t}]dt\\
    \ln{N(t)} &= \lambda t - \mu t + \frac{C_0}{\gamma}e^{-\gamma t} + B\\
    N(t) &= \frac{a}{e^{\frac{ C_0}{\gamma}}}e^{t(\lambda -\mu) + \frac{ C_0}{\gamma}e^{-\gamma t}}
\end{align*}

Using these values and derivations, we can derive a Kolmogorov Equation:

\begin{equation}
    \frac{dp_n(t)}{dt} = \lambda(n-1)p_{n-1}(t) + [\mu +  C_0e^{-\gamma t}](n+1)p_{n+1}(t) - [(\lambda + \mu) + C_0e^{-\gamma t}]np_n(t)
\end{equation}

Assume that we are given the initial condition $p_a(0) = 1$ with only one singular cell at the beginning of time in the cell population. Note that you can have different starting cell values, however it is classic to start with a singular cell body.

This is an example of the Quasi-Steady Stationary State Approximation, because for small $\gamma$, over short intervals, $C(t)$ changes slowly allowing us to treat the concentration of the treatment cells as approximately linear over these intervals.

Some notes about the terms we have appearing in the Kolmogorov Equation above:

\begin{itemize}
    \item \textbf{Birth rate $\lambda$}: The term $\lambda (n-1) p_{n-1}(t)$ represents the rate at which the population increases by one unit (birth). The coefficient $\lambda (n-1)$ indicates that the rate is proportional to the current population size minus one.
    \item \textbf{Death rate $\mu$}: The term $\mu(n+1) p_{n+1}(t)$ represents the rate at which the population decreases by one unit (death). The coefficient $\mu (n+1)$ indicates that the rate is proportional to the current population size plus one.
    \item \textbf{Treatment term $C_0e^{-\gamma t}(n+1)$}: This term represents a treatment or decay process where the rate is influenced by a treatment concentration that decays exponentially over time. $C_0$ is the initial concentration and $\gamma$ is the decay rate of the treatment.
    \item \textbf{Balance Term}: The term $-[(\lambda + \mu) +  C_0 e^{-\gamma t}]n p_n(t)$ ensures the probability is conserved and accounts for the transitions out of the state $n$.
\end{itemize}

\subsection{Probability Generating Function}

We can now start solving for the probability generating function, which we will define as $G(s, t)$.

\begin{align*}
\begin{split}
    \frac{dp_n(t)}{dt} &= \lambda(n-1) p_{n-1}(t) + [\mu + C_0e^{-\gamma t}](n+1)p_{n+1}(t) - [(\lambda + \mu) +  C_0e^{-\gamma t}]np_n(t)\\
    \sum_{n=0}^\infty s^n \frac{dp_n(t)}{dt} &= \sum s^n [\lambda(n-1) p_{n-1}(t) + [\mu +  C_0e^{-\gamma t}](n+1)p_{n+1}(t) - [(\lambda + \mu) +  C_0e^{-\gamma t}]np_n(t)]\\
    \implies &\sum s^n(\lambda(n-1)p_{n-1}(t)) + \sum s^n(\mu(n+1)p_{n+1}(t)) + \sum s^n ( C_0e^{-\gamma t}(n+1)p_{n+1}(t))-\\
    &- \sum s^n (\lambda + \mu)np_n(t) - \sum s^n( C_0 e^{-\gamma t})np_n(t)\\
    \implies &\lambda s^2 \sum s^{n-2}(n-1)p_{n-1}(t) + \mu \sum s^n(n+1)p_{n+1}(t) +  C_0e^{-\gamma t} \sum s^n(n+1)p_{n+1}(t) - \\
    &- (\lambda + \mu)s \sum s^{n-1}np_n(t) -  C_0e^{-\gamma t} s \sum s^{n-1}n p_n(t)\\
    \implies &\lambda s^2 \frac{d}{ds} \left[\sum s^{n-1} p_{n-1}(t)\right] + \mu \frac{d}{ds} \left[\sum s^{n+1} p_{n+1}(t)\right] +  C_0e^{-\gamma t} \frac{d}{ds} \left[\sum s^{n+1} p_{n+1}(t)\right] -\\ 
    &- (\lambda + \mu)s \frac{d}{ds} \left[\sum s^n p_n(t)\right] -  C_0e^{-\gamma t} s \frac{d}{ds} \left[\sum s^n p_n(t)\right]\\
    \implies \frac{\partial G(s, t)}{\partial t} &= (\lambda s^2 + \mu +  C_0e^{-\gamma t} - (\lambda + \mu)s -  C_0e^{-\gamma t} s) \frac{\partial G(s, t)}{\partial s}\\
    \frac{\partial G(s, t)}{\partial t} &= (\lambda s^2 - s(\lambda + \mu +  C_0e^{-\gamma t}) + (\mu +  C_0e^{-\gamma t}))\frac{\partial G(s, t)}{\partial s}
    \end{split}
\end{align*}

As this is a first order partial differential equation, we can solve it easily using Mathematica (code in Appendix \ref{appendix:PGF}
). The final result is as follows:\\

\begin{equation}
    G(s, t) = C_1 \left[ t - \frac{e^{\gamma t} \log{\left[C_0(s-1)+e^{\gamma t}(-\lambda^2 - \mu + s(\lambda + \mu))\right]}}{C_0 + e^{\gamma t}(\lambda + \mu)} \right]
\end{equation}

Where $C_1$ is an integration constant that can be solved for using the initial condition $G(s, 0) = s^a$.

To find \( C_1 \), we use the initial condition \( G(s, 0) = s^a \). Plugging \( t = 0 \) into the expression for \( G(s, t) \):

\[
G(s, 0) = C_1 \left[ 0 - \frac{e^{\gamma \cdot 0} \log{\left[C_0(s-1) + e^{\gamma \cdot 0}(-\lambda^2 - \mu + s(\lambda + \mu))\right]}}{C_0 + e^{\gamma \cdot 0}(\lambda + \mu)} \right]
\]

Simplifying this expression:

\[
G(s, 0) = C_1 \left[ - \frac{\log{\left[C_0(s-1) + (-\lambda^2 - \mu + s(\lambda + \mu))\right]}}{C_0 + \lambda + \mu} \right]
\]

Given \( G(s, 0) = s^a \), we can set this equal to the simplified expression above:

\[
s^a = C_1 \left[ - \frac{\log{\left[C_0(s-1) + (-\lambda^2 - \mu + s(\lambda + \mu))\right]}}{C_0 + \lambda + \mu} \right]
\]

To isolate \( C_1 \), we rearrange the equation:

\[
C_1 = \frac{s^a (C_0 + \lambda + \mu)}{- \log{\left[C_0(s-1) + (-\lambda^2 - \mu + s(\lambda + \mu))\right]}}
\]

Thus, the probability generating function \( G(s, t) \) can be written as:

\[
G(s, t) = \frac{s^a (C_0 + \lambda + \mu)}{- \log{\left[C_0(s-1) + (-\lambda^2 - \mu + s(\lambda + \mu))\right]}} \left[ t - \frac{e^{\gamma t} \log{\left[C_0(s-1) + e^{\gamma t}(-\lambda^2 - \mu + s(\lambda + \mu))\right]}}{C_0 + e^{\gamma t}(\lambda + \mu)} \right]
\]

Once we have the probability generating function \( G(s, t) \), the next step is to use it to derive the moments or the probabilities \( p_n(t) \). Here’s how we can proceed:

\subsection{First Moment (Mean)}

The first moment, or the mean of the distribution \( N(t) \), can be found by differentiating \( G(s, t) \) with respect to \( s \) and then setting \( s = 1 \).

\[
\mathbb{E}[X(t)] = \left. \frac{\partial G(s, t)}{\partial s} \right|_{s=1}
\]

Where $X(t)$ is treated as a random variable. Using Mathematica, code in the Appendix \ref{appendix:FirstMoment} we see that the mean value is:

\[
-\frac{a (C_0+\lambda +\mu ) \left(t-\frac{e^{\gamma  t} \log \left(\left(\lambda -\lambda ^2\right) e^{\gamma  t}\right)}{C_0+(\lambda +\mu ) e^{\gamma  t}}\right)}{\log \left(\lambda -\lambda ^2\right)}+\frac{C_0+\lambda +\mu }{\left(\lambda -\lambda ^2\right) \log \left(\lambda -\lambda ^2\right)}+\frac{(C_0+\lambda +\mu )^2 \left(t-\frac{e^{\gamma  t} \log \left(\left(\lambda -\lambda ^2\right) e^{\gamma  t}\right)}{C_0+(\lambda +\mu ) e^{\gamma  t}}\right)}{\left(\lambda -\lambda ^2\right) \log ^2\left(\lambda -\lambda ^2\right)}
\]

\subsection{Second Moment Calculation}

To compute the second moment, we need to evaluate the second derivative of the probability generating function \(G(s, t)\) with respect to \(s\) and then set \(s = 1\). 

\[
\textit{Var}[X(t)^2] = \left. \frac{\partial^2 G(s, t)}{\partial s^2} \right|_{s=1}
\]

Let's compute the second derivative of \(G(s, t)\) using the Mathematica code highlighted in \ref{appendix:SecondMoment}:

\begin{multline*}
    -2 a (C_0+\lambda +\mu ) \left(-\frac{(C_0+\lambda +\mu ) \left(t-\frac{e^{\gamma  t} \log \left(\left(\lambda -\lambda ^2\right) e^{\gamma  t}\right)}{C_0+(\lambda +\mu ) e^{\gamma  t}}\right)}{\left(\lambda -\lambda ^2\right) \log ^2\left(\lambda -\lambda ^2\right)}-\frac{1}{\left(\lambda -\lambda ^2\right) \log \left(\lambda -\lambda ^2\right)}\right)-\\
    -\frac{(a-1) a (C_0+\lambda +\mu ) \left(t-\frac{e^{\gamma  t} \log \left(\left(\lambda -\lambda ^2\right) e^{\gamma  t}\right)}{C_0+(\lambda +\mu ) e^{\gamma  t}}\right)}{\log \left(\lambda -\lambda ^2\right)}-\\
    -(C_0+\lambda +\mu ) \Bigg(\frac{2 (C_0+\lambda +\mu )}{\left(\lambda -\lambda ^2\right)^2 \log ^2\left(\lambda -\lambda ^2\right)}+\left(\frac{2 (C_0+\lambda +\mu )^2}{\left(\lambda -\lambda ^2\right)^2 \log ^3\left(\lambda -\lambda ^2\right)}+\frac{(C_0+\lambda +\mu )^2}{\left(\lambda -\lambda ^2\right)^2 \log ^2\left(\lambda -\lambda ^2\right)}\right)\cdot\\
    \cdot\left(t-\frac{e^{\gamma  t} \log \left(\left(\lambda -\lambda ^2\right) e^{\gamma  t}\right)}{C_0+(\lambda +\mu ) e^{\gamma  t}}\right)+\frac{e^{\gamma  (-t)} \left(C_0+(\lambda +\mu ) e^{\gamma  t}\right)}{\left(\lambda -\lambda ^2\right)^2 \log \left(\lambda -\lambda ^2\right)}\Bigg)
\end{multline*}

One thing to note here, finding a correlation between the First and Second moment proved to be too hectic and no analysis relating the two has been done. Further work on this can be done in related or future work.

\subsection{Finding the Probabilities \( p_n(t) \)}

Due to the complexity of finding a pattern and multiple derivatives that are needed in finding the probability formula $p_n(t)$, we pursue a more numerical approach and numerically solve the problem. The code is included in the Appendix \ref{appendix:MATLAB Probability}, and the motivation for this code was derived from \href{https://github.com/kmamis/stochastic_model_colonic_crypts/blob/main/matlab_code_for_fig2.m}{here} \cite{GIT}.

Note in the plots below, time is defined in minutes.

The result is:

\begin{center}
    \includegraphics[width=0.75\textwidth]{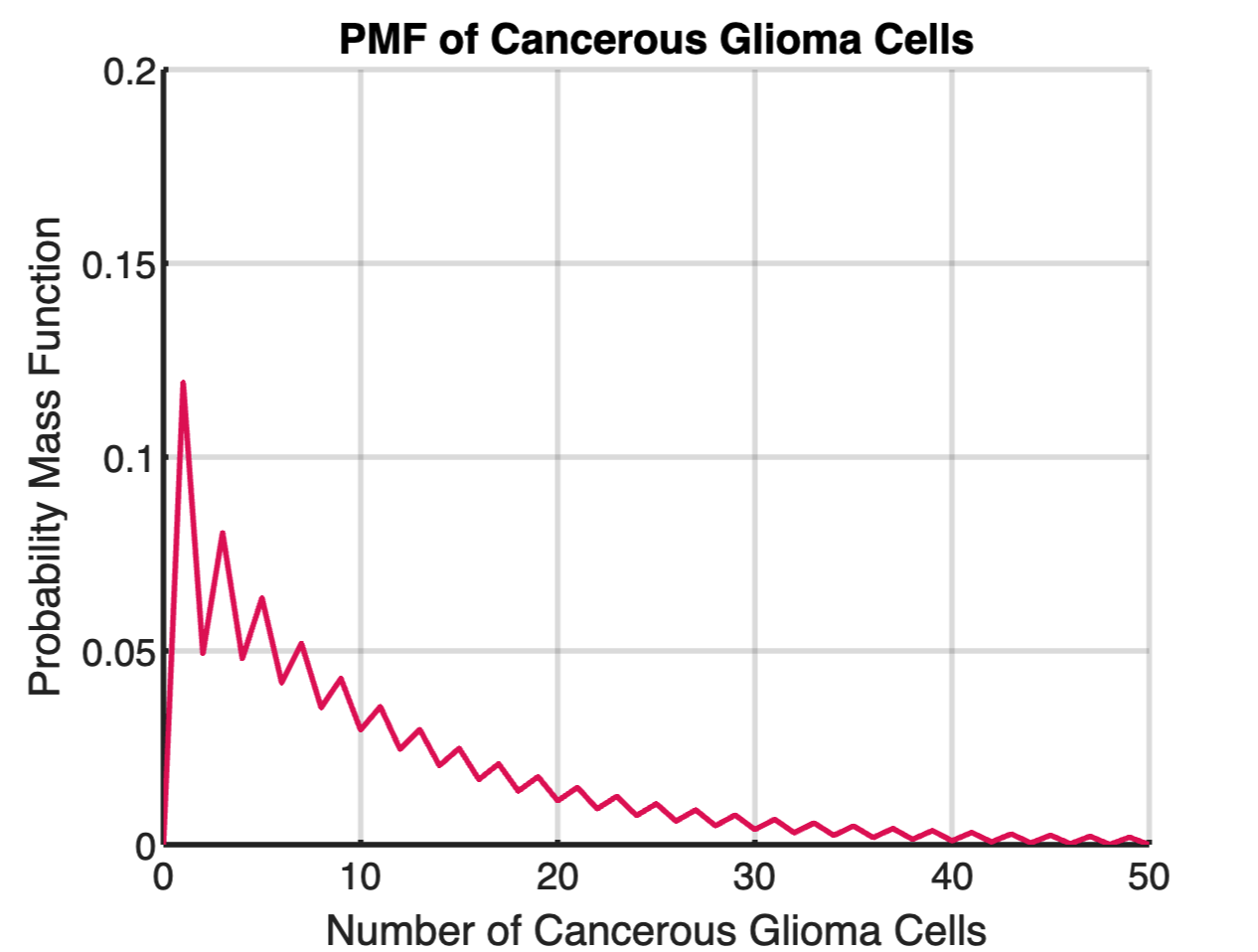}
\end{center}

\subsubsection*{PMF of Cancerous Glioma Cells}
 The first plot shows the probability mass function (PMF) of the number of cancerous glioma cells at a specific time (t = 10 minutes). The graph indicates that the most probable number of cancerous cells is relatively low, with the probability decreasing as the number of cells increases. This suggests that, at this time point, smaller cell populations are more likely compared to larger ones, indicating a relatively early stage of cancer cell proliferation.

Here is another plot showing the Mean number of Glioma Cancerous cells over Time.

\begin{center}
    \includegraphics[width=0.75\textwidth]{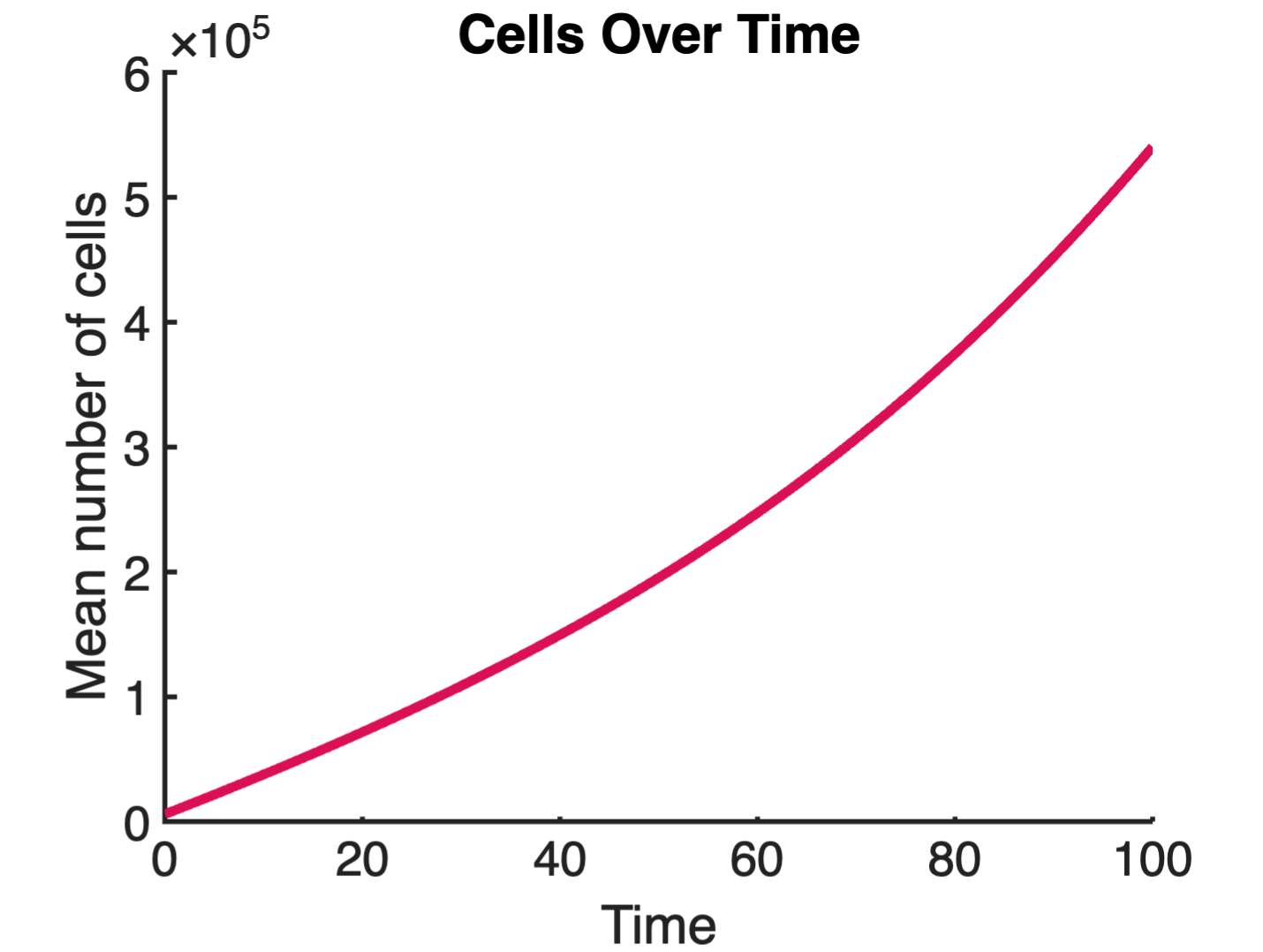}
\end{center}

\subsubsection*{Mean Number of Glioma Cancerous Cells Over Time}
The second plot displays the mean number of glioma cancerous cells over time, demonstrating an exponential growth pattern. The curve indicates that the average number of cancerous cells increases rapidly as time progresses. This highlights the aggressive nature of glioma cell proliferation, emphasizing the need for timely and effective treatment interventions to manage the cancer growth.

\section{Visualization and Simulation}

\subsection{Finding Parameter Values} \label{parameter values}

For the birth rate $\lambda$ and the death rate $\mu$, these will need to be approximated via simulations to see what mix of those parameters fit our expectations the best.\\

However for $a$, $C_0$, and $\gamma$, we can find values for those. According to the paper \cite{Glioma}, under the dose-response section of the Cytochalasin D graphics section, it was found that the IC50 value was $15.43$ micro-moles ($\mu$M units), meaning that the population would be cut in half after the 1 hour experiment. We can also observe this phenomena approximately from the graphic given in the paper \cite{Glioma} included here:\\

\begin{center}
    \includegraphics[]{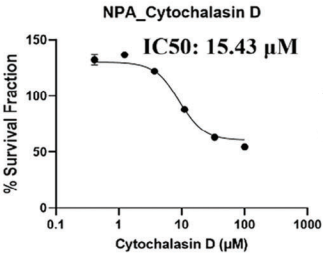}
\end{center}

Thus, we can conclude that $C_0=15.43\mu$M. We can adjust this parameter by changing $\beta=1$ (the treatment scaling factor) to $\beta=\frac{1}{100}$ to allow for the ODE model (\ref{Cytochalasin D Treatment}) to be interpretable in a graphical sense, so can be $C_0=0.1543\mu$M. Subsequently, it was also done for the simulations runs to keep our parameters consistent. Now for $a$, the initial population, the paper offers two initial populations, one of low density cell count of $1
\times10^5$ NPA cells, and one of high density cell count of $2\times10^5$ NPA cells. At low density, no oncostreams were observed, whereas high density conditions facilitated robust oncostream formation. So it is safe to assume that our initial population $a=2\times10^5$ NPA cells.\\

For $\gamma$, given the half-life time of Natural Cytochalasin D, as mentioned in the drug background information in section \ref{Cytochalasin D Treatment}, we can calculate this decay rate. Given that the half life is 10 minutes we can calculate $\gamma$ as:

$$\displaystyle{\text{Given that: } N_0\left(\dfrac{1}{2}\right)^{\dfrac{t}{t_{\frac{1}{2}}}}=N_0e^{-\gamma t}\text{ are both exponential decay equations}}$$
$$\displaystyle{t_{\frac{1}{2}} \text{ represents half-life, we can solve for }\gamma}$$

\begin{equation}
    \begin{split}
        \left(\dfrac{1}{2}\right)^{\dfrac{t}{t_{\frac{1}{2}}}} &= e^{-\gamma t} \\
        \ln\left(\left(\dfrac{1}{2}\right)^{\dfrac{t}{t_{\frac{1}{2}}}}\right) &= \ln\left(e^{-\gamma t}\right)\\
        \frac{t}{t_{\frac{1}{2}}}\ln\left(\frac{1}{2}\right) &= -\gamma t\\
        -\frac{t}{t_{\frac{1}{2}}}\ln(2) &= -\gamma t\\
        \frac{\ln(2)}{t_{\frac{1}{2}}} &= \gamma\\
    \end{split}
\end{equation}

From here, we can explicitly calculate $\gamma_1$ as:
$$\displaystyle{\gamma_1 = \frac{\ln(2)}{10\text{ minutes}}\approx\frac{0.0693}{\text{minute}}}$$\\

The same process can be followed for Liposomal Cytochalasin Treatment to find that:
$$\displaystyle{\gamma_2 = \frac{\ln(2)}{10\text{ minutes}}\approx\frac{0.0029}{\text{minute}}}$$\\

So to reiterate, our known parameters for Natural Cytochalasin Treatment are:
\begin{align*}
    a &= 2\times10^5\text{ NPA cells}\\
    C_0 &= 0.1543\text{ micro-moles }(\mu\text{M})\\
    \gamma_1 &= \frac{0.0693}{\text{minute}}
\end{align*}

And the for Liposomal Cytochalasin Treatment, the only difference is:
\begin{align*}
    \gamma_2 &= \frac{0.0029}{\text{minute}}
\end{align*}

\subsection{Calculating Probability Distribution of the Time Until the Next Event for Simulations}

For this process, we can use a known relation and then solve for $f_T(t)$, the probability density function of the time, $T$, the time until the next event.\\

We know that:
$$\displaystyle{\mathbb{P}(T>t+\Delta t)=\mathbb{P}(T>t)\mathbb{P}_{\text{no event during $\Delta t$}}}$$\\

Which can be written as:
$$\displaystyle{\mathbb{P}(T>t+\Delta t)=\biggr(1-(\lambda+\mu+C(t))n\Delta t\biggl)\mathbb{P}(T>t)}$$\\
$$\displaystyle{\mathbb{P}(T>t+\Delta t)=\biggr(1-(\lambda+\mu+C_0e^{-\gamma t})n\Delta t\biggl)\mathbb{P}(T>t)}$$\\
$$\displaystyle{\mathbb{P}(T>t+\Delta t)-\mathbb{P}(T>t)=-(\lambda+\mu+C_0e^{-\gamma t})n\Delta t\mathbb{P}(T>t)}$$\\
$$\displaystyle{\frac{\mathbb{P}(T>t+\Delta t)-\mathbb{P}(T>t)}{\Delta t}=\frac{-(\lambda+\mu+C_0e^{-\gamma t})n\Delta t\mathbb{P}(T>t)}{\Delta t}}$$\\
This is the form of the derivative, thus:
$$\displaystyle{\frac{d{P}(T>t)}{dt}=-(\lambda+\mu+C_0e^{-\gamma t})n\mathbb{P}(T>t)}$$\\
And we can easily solve this ODE, with $A$ being an integration constant:
$$\displaystyle{\mathbb{P}(T>t)=A\exp\left[-(\lambda+\mu)nt+\dfrac{C_0ne^{-\gamma t}}{\gamma}\right], A\in\mathbb{R}}$$\\
We can solve for $A$ by using the fact that $\mathbb{P}(T>0) = 1$:
$$\displaystyle{\mathbb{P}(T>0)=1=A\exp\left[0+\dfrac{C_0ne^{0}}{\gamma}\right]}$$
$$\displaystyle{1=A\exp\left[\dfrac{C_0n}{\gamma}\right]}$$
$$\displaystyle{e^{-\dfrac{C_0n}{\gamma}}=A}$$
Then we can express this as the cumulative distribution function of $T$:
$$\displaystyle{F_T(t)=1-\mathbb{P}(T>t)=1-e^{-\dfrac{C_0n}{\gamma}}\exp\left[-(\lambda+\mu)nt+\dfrac{C_0ne^{-\gamma t}}{\gamma}\right]}$$
$$\displaystyle{F_T(t)=1-\mathbb{P}(T>t)=1-\exp\left[-(\lambda+\mu)nt+\dfrac{C_0n(e^{-\gamma t}-1)}{\gamma}\right]}$$
And then we can get the probability density function by taking the derivative of $F_T(t)$: 
$$\displaystyle{f_T(t)=n\left(C_0e^{-\gamma t}+\lambda+\mu\right)\exp\left[-(\lambda+\mu)nt+\frac{C_0n(e^{-\gamma t}-1)}{\gamma}\right]}$$
As a sanity check, we can confirm that the probability density function is legitimate by integrating it. Using Mathematica (code in Appendix \ref{appendix:PDF}) we can see that the integral produces 1, meaning it is a legitimate probability density function.
$$\displaystyle{\int_0^{\infty}f_T(t)dt=\int_0^{\infty}n\left(C_0e^{-\gamma t}+\lambda+\mu\right)\exp\left[-(\lambda+\mu)nt+\frac{C_0n(e^{-\gamma t}-1)}{\gamma}\right]dt = 1}$$

Here are graphs representing the distribution for both Natural and Liposomal Cytochalasin D. The y-axis was scaled down by a factor of 100 to show probabilities a decimal to better represent what the probability is. A population of $10$ was used to allow better representation of and visualization of the PDF curve. Both of these plots were generated using code that can be found in Appendix \ref{appendix:MATLAB PDF graphs}.

\begin{center}
    \includegraphics[scale = 0.5]{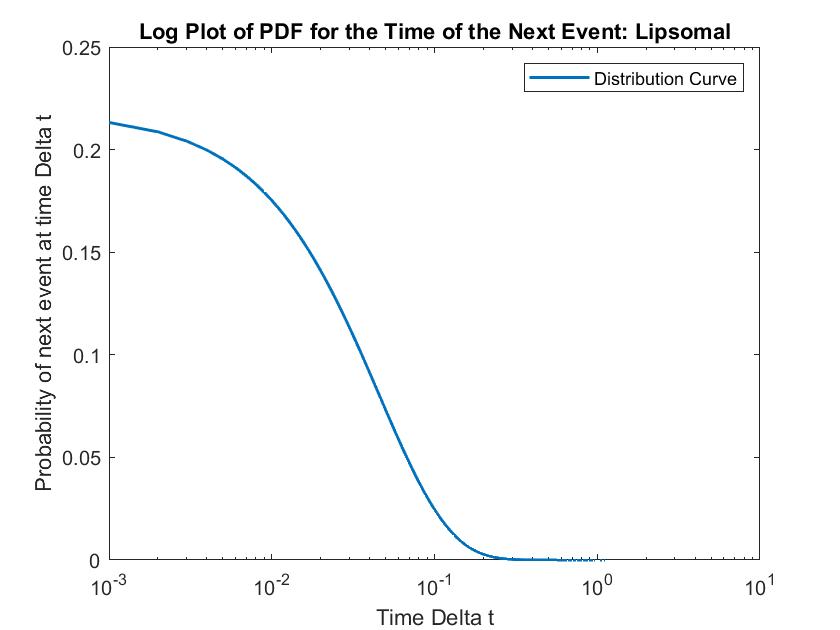}\\
    \includegraphics[scale = 0.5]{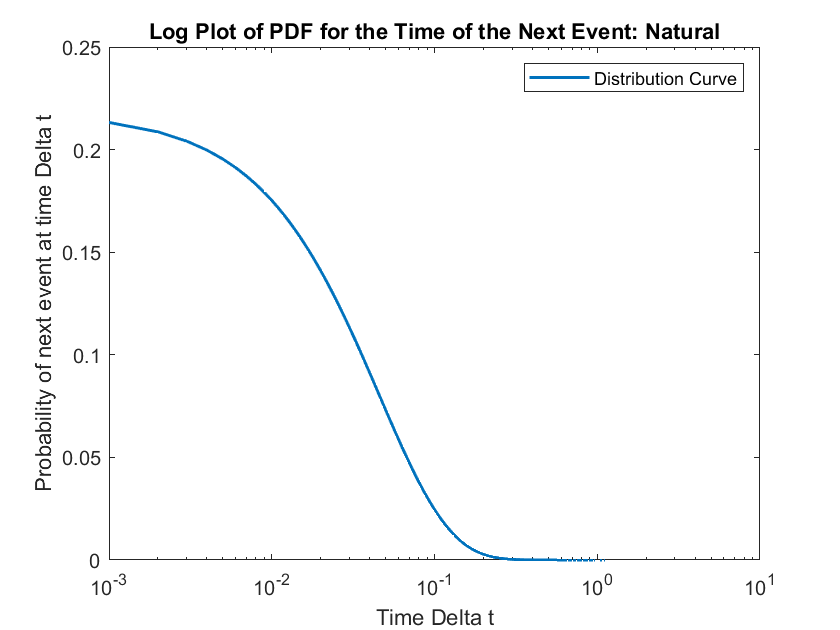}
\end{center}

From these plots, it is easy to observe that the probability of the next event happening is highest when $0<\Delta t << 1$. As a result of this, births and deaths will occur extremely frequently with very little time in between each birth or death.\\

Simulation results of 10 different runs yielded these graphs (using MATLAB code found in Appendix \ref{appendix:MATLAB PDF graphs}).\\

\begin{figure}[h]
    \centering
    \includegraphics[scale = 0.4]{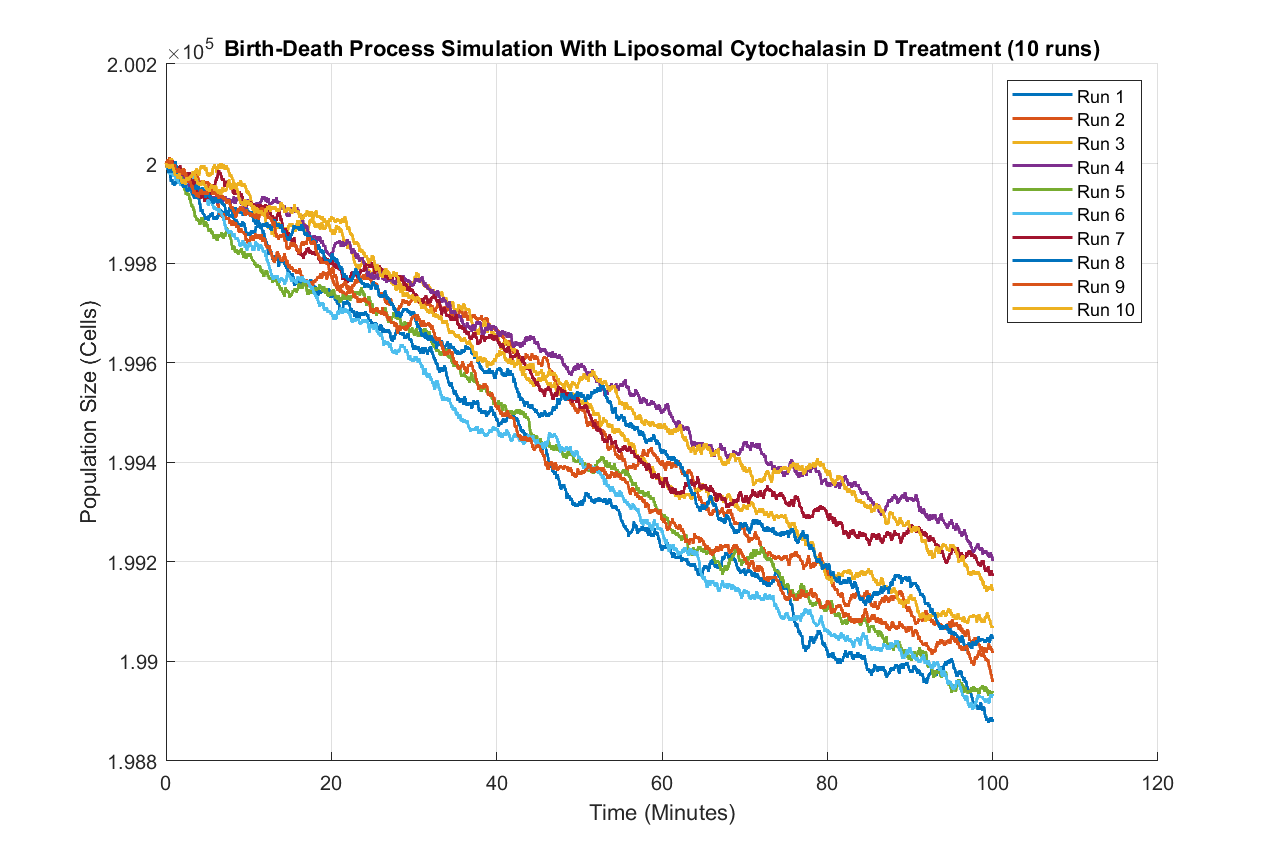}
\end{figure}

In this Birth-Death simulation with Liposomal Cytochalasin D Treatment, we can see that on average every simulation run has the same downward trend and actually reduces the population size as time passes. This is very good because this indicates that the treatment is working properly and shows promising results for controlling and reducing the Cell Population.

\begin{figure}
    \centering
    \includegraphics[scale = 0.4]{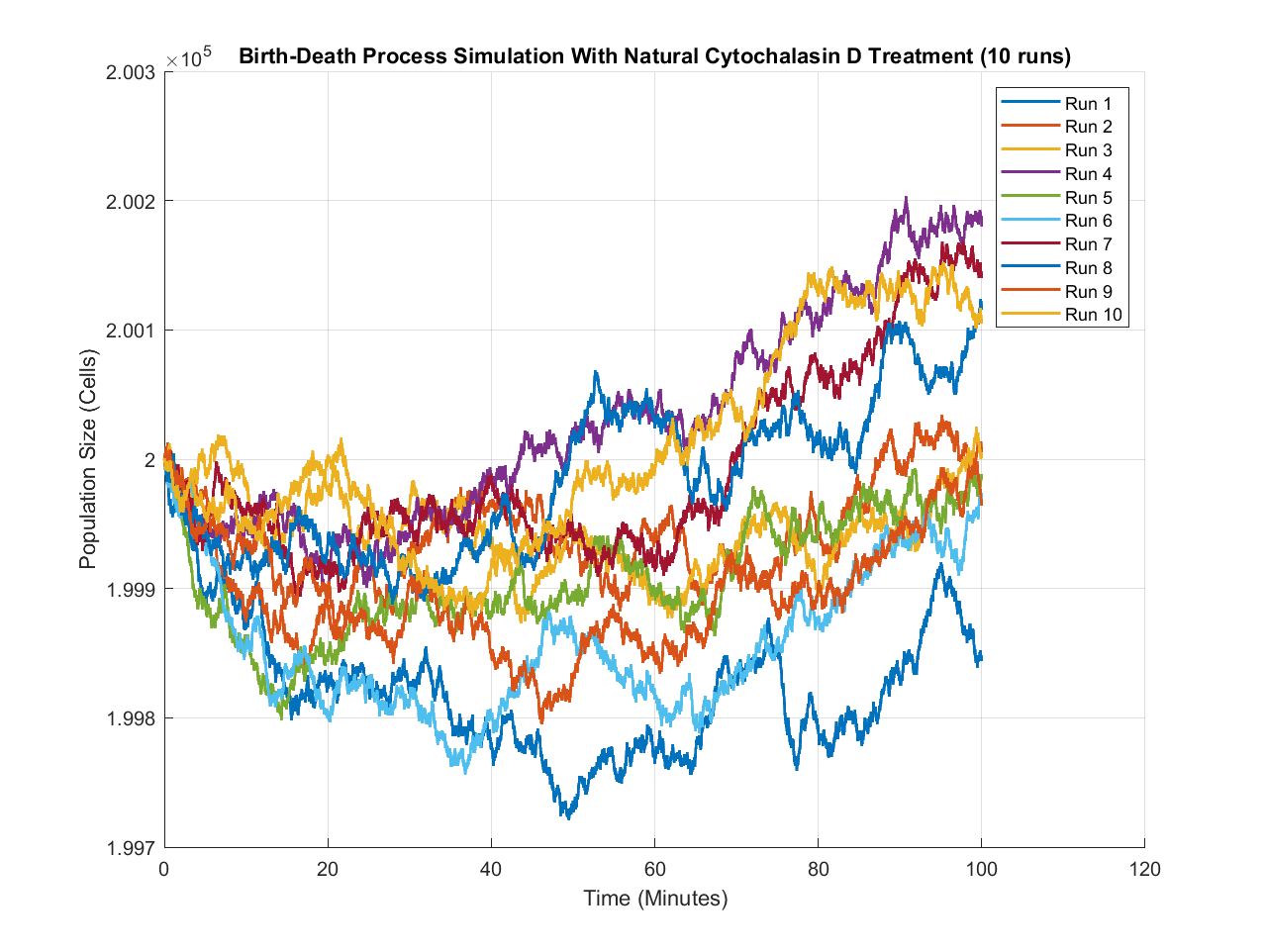}
\end{figure}

In this Birth-Death Process simulation with Natural Cytochalasin D Treatment, we can see that it doesn't actually work as much as the previous type of treatment. With the Population growing exponentially in some runs, and the overall trend still showing that the population grows.

\newpage

\section{Analysis}

ODE Model using parameters stated in section \ref{parameter values} using MATLAB code found in Appendix \ref{appendix:MATLAB ODE graph}.

\begin{center}
    \includegraphics[scale=0.75]{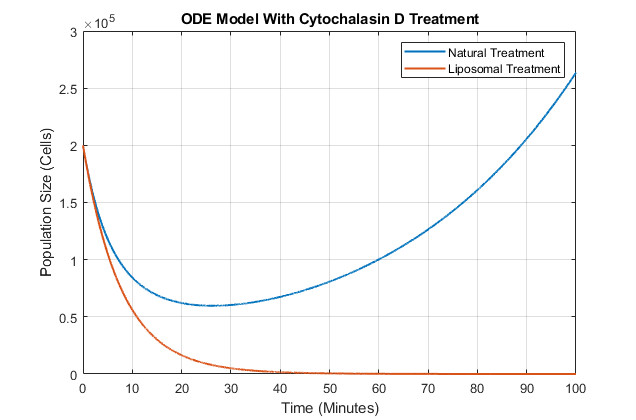}
\end{center}

The main goal this paper is to be able to somewhat match up with the ODE model with the Stochastic model. In this case, we were somewhat unsuccessful, however an important observation is that the general behavior and trend of the two models are comparable. The shape of graphs for both treatment types follow similar behavioral paths, with Natural Cytochalasin Treatment decaying the population, and then seeing a rebound. And with Liposomal Cytochalain Treatment, we see a continued decay of the population over the entire course of the experiment, which makes sense because the Liposomal Treatment half-life is 24 times longer than that the Natural Treatment.\\

A possible solution to the issues we faced trying to model this would fixing the unit conversions between different variables. Some unit conversions were assumed to be viable, and to allow for simplicity in the first steps of creating a model, and then were continued to be assumed to see a result that was easily interpretable. Obviously it is an issue that the average population from the simulation runs isn't similar to the population from the ODE model. But because of the similar behavior between the ODE and Stochastic models, this issue could be fixed by possibly changing some coefficients on the models to have the exact correct unit conversions which would hopefully align the graphs in a way where the cell populations are much more comparable than what we have currently.

\newpage

\section{Conclusion}

In conclusion, this research provides significant insights into the dynamics of glioma progression through the study of oncostreams, utilizing advanced ex-vivo 3D explants and in-vivo intravital imaging techniques. By focusing on the critical role of oncostreams and modeling the cell populations within these structures, particularly GFP+ NPA cells, this study has successfully established a direct correlation between oncostream density and cancer severity. The use of a simple Birth-Death process and Kolmogorov equations to analyze cell population dynamics and treatment effects has provided a robust quantitative framework to predict changes over time.

The findings from this study highlight the intricate relationship between oncostream density and glioma aggressiveness. The investigation into various treatment methods, particularly the use of Cytochalasin D, has shown promise in disrupting oncostreams and reducing glioma invasiveness. This disruption of oncostreams presents a potential therapeutic strategy, offering new avenues for glioma treatment that could significantly impact patient outcomes.

The analysis section reveals that while there was some difficulty in achieving an exact match between the ODE model and the Stochastic model, the general behavior and trends of the two models were comparable. This suggests that the foundational approach is sound, though further refinement in unit conversions and coefficients is necessary to enhance accuracy. The similar behavioral patterns observed between the ODE and Stochastic models, despite these discrepancies, indicate a solid basis for future work in this area.

The integration of Cytochalasin D into the model has demonstrated its potential in reducing glioma invasiveness by disrupting oncostreams. This aligns with the hypothesis that targeting oncostreams can significantly impact glioma dynamics and progression. The analysis of treatment effects over time, as illustrated in the ODE model, supports the potential for Cytochalasin D to serve as a valuable component of glioma therapy.

Future work should focus on refining the models by addressing unit conversion issues and incorporating additional modulators such as morphological attributes and neurotransmitter exposure. This expansion would provide a more comprehensive understanding of glioma behavior and the factors influencing its progression. The development of more accurate models will enhance the predictive power and applicability of this research, leading to better-informed therapeutic strategies.

In summary, this research contributes to the growing body of knowledge on glioma progression by elucidating the role of oncostreams and exploring potential treatment strategies. The findings underscore the importance of targeting oncostreams to disrupt glioma invasiveness and highlight the therapeutic potential of agents like Cytochalasin D. By advancing our understanding of glioma dynamics and identifying effective interventions, this study paves the way for improved cancer management strategies, offering hope for better outcomes for patients suffering from this devastating disease.

\newpage

\bibliography{references}
\bibliographystyle{plain}

\section{Simple Birth-Death process derivation}
\label{appendix:derivation}

Starting with the forward Kolmogorov equation:\\
    $$\displaystyle{\frac{dp_n(t)}{dt}=\lambda(n-1)p_{n-1}(t) +\mu(n+1)p_{n+1}(t)-(\lambda+\mu)np_n(t)}$$\\
    We want to get in the form of a probability generating function (PGF), $$G(s, t) = \sum_{n=0}^\infty p_n(t)s^n$$ so we multiply each term by $s^n$ and sum every term from $n=0$ to $\infty$, and then following some algebraic manipulations:\\
    $$\displaystyle{\sum_{n=0}^{\infty}s^n\frac{dp_n(t)}{dt}=\sum_{n=0}^{\infty}s^n\lambda(n-1)p_{n-1}(t) +\sum_{n=0}^{\infty}s^n\mu(n+1)p_{n+1}(t)-\sum_{n=0}^{\infty}s^n(\lambda+\mu)np_n(t)}$$\\
    $$\displaystyle{\frac{d}{dt}\sum_{n=0}^{\infty}s^np_n(t)=\lambda s^2\sum_{n=0}^{\infty}s^{n-2}(n-1)p_{n-1}(t) +\mu\sum_{n=0}^{\infty}s^n(n+1)p_{n+1}(t)-(\lambda+\mu)s\sum_{n=0}^{\infty}s^{n-1}np_n(t)}$$\\
    $$\displaystyle{\frac{d}{dt}\sum_{n=0}^{\infty}s^np_n(t)=\lambda s^2\frac{d}{ds}\left[\sum_{n=0}^{\infty}s^{n-1}p_{n-1}(t)\right] + \mu\frac{d}{ds}\left[\sum_{n=0}^{\infty}s^{n+1}p_{n+1}(t)\right]-(\lambda+\mu)s\frac{d}{ds}\left[\sum_{n=0}^{\infty}s^np_n(t)\right]}$$\\
    $$\displaystyle{\frac{\partial G(s,t)}{\partial t}= \left(\lambda s^2-\left(\lambda +\mu \right)s+ \mu \right)\frac{\partial G(s,t)}{\partial s}}$$\\

    Thus, we have found the PDE for the probability generating function.\\

    Next we need the characteristic equations:
     $$\displaystyle{\frac{dt}{1}=-\frac{ds}{\lambda s^2-(\lambda +\mu)s +\mu}=\frac{dG}{0}}$$
 
     From this, we can easily see that $G$ is a constant, so we only need to solve:
     $$\displaystyle{\frac{dt}{1}=-\frac{ds}{\lambda s^2-(\lambda +\mu)s +\mu}}$$
     First we will solve for $\lambda=\mu$:
     $$\displaystyle{\frac{dt}{1}=-\frac{ds}{\lambda s^2-2\lambda s +\lambda}}$$
     $$\displaystyle{\int dt = \int\frac{-ds}{\lambda s^2-2\lambda s +\lambda}=\frac{1}{\lambda}\int\frac{-ds}{(s-1)^2}}$$
     Using u-substitution:
     $$\displaystyle{t=\frac{1}{\lambda(s-1)}+C},\text{ where } C \text{ is an integration constant}$$
     $$\displaystyle{C=t-\frac{1}{\lambda(s-1)}}$$
     We know that $G$ is a constant, so for some function $\Psi$, we can do as follows:
     $$\displaystyle{G(s,t)=\Psi\left(t-\frac{1}{\lambda(s-1)}\right)}$$
     To determine $\Psi$, we need to use initial condition of $p_a(0)=1$ and we know that:
     $$\displaystyle{G(s,0)=s^a=\Psi\left(-\frac{1}{\lambda(s-1)}\right)}$$
     Now if we let $u=-\dfrac{1}{\lambda(s-1)}$, it is easy to see that $s=1-\dfrac{1}{u\lambda}$, thus we can do:
     $$\displaystyle{\Psi\left(u\right)=\left(1-\frac{1}{u\lambda}\right)^a}$$
     Now we can let $u=t-\dfrac{1}{\lambda(s-1)}$ and:
     $$\displaystyle{G(s,t)=\left(1-\frac{1}{\lambda\left(t-\frac{1}{\lambda(s-1)}\right)}\right)^a}$$
     Skipping over simple algebraic manipulations, we arrive here:
     $$\displaystyle{G(s,t)=\left(1-\frac{s-1}{\lambda t(s-1)-1}\right)^a}$$
     And then here:
     $$\displaystyle{G(s,t)=\left(\frac{1-(\lambda t-1)(s-1)}{1-\lambda t(s-1)}\right)^a}$$\\

     Now for $\lambda \ne \mu$, we can follow the derivations in the Bailey textbook \cite{Bailey}:\\
     
     From this point:\\
     $$\displaystyle{\frac{dt}{1}=-\frac{ds}{\lambda s^2-(\lambda +\mu)s +\mu}}$$\\
     $$\displaystyle{\int dt = \int\frac{-ds}{\lambda s^2-(\lambda + \mu)s +\mu}}$$
     Which results in:\\
     $$\displaystyle{t=\frac{\ln\left(\left|{\lambda}s-{\mu}\right|\right)-\ln\left(\left|s-1\right|\right)}{{\lambda}-{\mu}} + C}$$
     With $C$ as an integration constant, we can do some algebraic manipulations to arrive at:\\
     $$\displaystyle{t=-\frac{\ln(s-1)-\ln(\lambda s-\mu)}{\lambda-\mu} + C}$$\\
     $$\displaystyle{t=-\frac{1}{\lambda-\mu}\ln\left(\frac{s-1}{\lambda s - \mu}\right) + C}$$\\
     $$\displaystyle{C=t+\frac{1}{\lambda-\mu}\ln\left(\frac{s-1}{\lambda s - \mu}\right)}$$\\
     $$\displaystyle{C=t+\frac{1}{\lambda-\mu}\ln\left(\frac{s-1}{\lambda s - \mu}\right)}$$\\
     $$\displaystyle{(\lambda-\mu)C=(\lambda-\mu)t+\ln\left(\frac{s-1}{\lambda s - \mu}\right)}$$\\
     $$\displaystyle{e^{(\lambda-\mu)C}=e^{(\lambda-\mu)t+\ln\left(\frac{s-1}{\lambda s - \mu}\right)}}$$\\
     We can redefine the constant $C$ as $C=e^{(\lambda-\mu)C}$, as both $\lambda$ and $\mu$ are constants.\\
     $$\displaystyle{C=e^{(\lambda-\mu)t}e^{\ln\left(\frac{s-1}{\lambda s - \mu}\right)}}$$\\
     $$\displaystyle{C=\frac{(s-1)e^{(\lambda-\mu)t}}{\lambda s - \mu}}$$\\
     We know that $G$ is a constant, so for some function $\Psi$, we can do as follows:\\
     $$\displaystyle{G(s,t)=\Psi\left(\frac{(s-1)e^{(\lambda-\mu)t}}{\lambda s - \mu}\right)}$$\\
     To determine $\Psi$, we need to use initial condition of $p_a(0)=1$ and we know that:\\
     $$\displaystyle{G(s,0)=s^a=\Psi\left(\frac{s-1}{\lambda s - \mu}\right)}$$\\
     
     Now if we let $u=\dfrac{s-1}{\lambda s-\mu}$, it is easy to see that $s=\dfrac{\mu u-1}{\lambda u-1}$, thus we can do:\\
     
     $$\displaystyle{\Psi\left(u\right)=\left(\frac{\mu u-1}{\lambda u-1}\right)^a}$$\\
     Now we can let $u=\dfrac{(s-1)e^{(\lambda-\mu)t}}{\lambda s - \mu}$ and:\\
     $$\displaystyle{G(s,t)=\left(\frac{\mu \omega(s,t)-1}{\lambda \omega(s,t)-1}\right)^a \text{, where } \omega(s,t) = \dfrac{(s-1)e^{(\lambda-\mu)t}}{\lambda s - \mu}}$$\\

     Now we can solve explicitly for $p_n(t)$ for the special case of $a=1$, and find its exact distribution. Solving for $\lambda = \mu$ first:\\

     We know that $a=1$ so:\\
     $$\displaystyle{G(s,t)=\frac{1-(\lambda t-1)(s-1)}{1-\lambda t(s-1)}}$$\\

     We also know the formula:\\
     $$\displaystyle{p_n(t)=\frac{1}{n!}\cdot\frac{\partial^nG(s,t)}{\partial s^n}\bigg|_{s=0}}$$\\
     
     So the first derivative is:\\
     $$\dfrac{1}{\left(t{\lambda}s-t{\lambda}-1\right)^2}\bigg|_{s=0}=\dfrac{1}{(-1-\lambda t)^2}=\dfrac{1}{(-(1+\lambda t))^2}=\dfrac{1}{(1+\lambda t)^2}$$\\
     
     The second derivative:\\
     $$-\dfrac{2t{\lambda}}{\left(t{\lambda}s-t{\lambda}-1\right)^3}\bigg|_{s=0}=-\frac{2\lambda t}{(-(1+\lambda t))^3}=\frac{2\lambda t}{(1+\lambda t)^3}$$\\
     
     The third derivative:\\
     $$\dfrac{6t^2{\lambda}^2}{\left(t{\lambda}s-t{\lambda}-1\right)^4}\bigg|_{s=0}=\frac{6(\lambda t)^2}{(1+\lambda t)^4}$$\\
     
     The fourth derivative:\\
     $$-\dfrac{24t^3{\lambda}^3}{\left(t{\lambda}s-t{\lambda}-1\right)^5}\bigg|_{s=0}=\frac{24(\lambda t)^3}{(1+\lambda t)^5}$$\\
     
     We can see a pattern emerging, which can be written and solved for $p_n(t)$ as done below:\\
     $$\displaystyle{p_n(t)=\frac{1}{n!}\cdot\frac{n!(\lambda t)^{n-1}}{(1+\lambda t)^{n+1}}=\frac{(\lambda t)^{n-1}}{(1+\lambda t)^{n+1}}\text{, }n\ge1}$$
     $$\displaystyle{p_0(t)=\frac{\lambda t}{1 + \lambda t}}$$\\

     We can follow a similar derivation process for $\lambda\ne\mu$ for $a=1$, which results in:
     $$\displaystyle{p_n(t)=(1-\alpha)(1-\beta)\beta^{n-1}\text{, \space }\alpha=\frac{\mu(e^{(\lambda-\mu)t}-1)}{\lambda e^{(\lambda-\mu)t}-\mu}\text{, \space }\beta=\frac{\lambda(e^{(\lambda-\mu)t}-1)}{\lambda e^{(\lambda-\mu)t}-\mu}\text{, \space }n\ge1}$$
     $$\displaystyle{p_0(t)=\alpha}$$\\

     For results regarding $p_n(t)$ for $a>1$, a similar method can be used to find a much more complex answer, but one that is not necessary for the purposes of this paper.
     
\section{Mathematica Code for PGF}
\label{appendix:PGF}

\begin{lstlisting}[language=Mathematica, breaklines]
    pde = D[G[s, t], t] - ([Lambda]^2 - s ([Lambda] + [Mu] + C Exp[-t[Gamma]]) + [Mu] + 
       C Exp[-t[Gamma]]) D[G[s, t], s] == 0;
       sol = DSolve[pde, G[s, t], {s, t}]
\end{lstlisting}

\section{Mathematica Code for calculating First Moment of PGF}
\label{appendix:FirstMoment}

\begin{lstlisting}[language=Mathematica, breaklines]
    G[s_, t_, a_, C0_, lambda_, mu_, 
  gamma_] := (s^a*(C0 + lambda + mu))/(-Log[
      C0*(s - 1) + (-lambda^2 - mu + s*(lambda + mu))])*(t - (Exp[
        gamma*t]*
       Log[C0*(s - 1) + 
         Exp[gamma*t]*(-lambda^2 - mu + s*(lambda + mu))])/(C0 + 
       Exp[gamma*t]*(lambda + mu)))
partialDerivative = D[G[s, t, a, C0, lambda, mu, gamma], s]
partialDerivativeEvaluated = partialDerivative /. s -> 1
\end{lstlisting}

\section{Mathematica Code for calculating Second Moment of PGF}
\label{appendix:SecondMoment}

\begin{lstlisting}[language=Mathematica, breaklines]
    G[s_, t_, a_, C0_, lambda_, mu_, 
  gamma_] := (s^a*(C0 + lambda + mu))/(-Log[
      C0*(s - 1) + (-lambda^2 - mu + s*(lambda + mu))])*(t - (Exp[
        gamma*t]*
       Log[C0*(s - 1) + 
         Exp[gamma*t]*(-lambda^2 - mu + s*(lambda + mu))])/(C0 + 
       Exp[gamma*t]*(lambda + mu)))

secondPartialDerivative = D[G[s, t, a, C0, lambda, mu, gamma], {s, 2}]

secondPartialDerivativeEvaluated = secondPartialDerivative /. s -> 1
\end{lstlisting}

\section{Mathematica Code for PDF of Time of Next Event}
\label{appendix:PDF}

\begin{lstlisting}[language=Mathematica, breaklines]
    Integrate[
 n*(C/E^([Gamma]*t) + [Lambda] + [Mu])*
  E^((-([Lambda] + [Mu]))*n*
      t + (C*n*(-1 + E^((-[Gamma])*t)))/[Gamma]), {t, 0, Infinity}]
\end{lstlisting}

\section{MATLAB Code for PGF Graphs}
\label{appendix:MATLAB Probability}

\begin{lstlisting}[language = MATLAB, breaklines]
    clear all

% Model parameters
lambda = 1/30; % Birth rate (per hour)
mu = 1/(3.5*24); % Death rate (per hour)
gamma_treatment = 1/24; % Decay rate of treatment (per hour)
C0 = 10; % Initial guess for balance term

% Initial values for solving probabilities
n0 = 18;
ntot = 2392.10;
r = 1/(2.5*24); % (per hour)

% Calculation of TA differentiation rate d
syms y s t
d = solve((1 + r/(y - lambda) + r * y / (mu * (y - lambda))) * n0 == ntot, y);
d = double(d);

% Other auxiliary quantities
a = (d - lambda) / mu;

% Define the provided PGF
G = @(s, t) (s.^a * (C0 + lambda + mu)) ./ (-log(C0 * (s-1) + (lambda^2 - mu + s * (lambda + mu)))) .* ...
    (t - exp(gamma_treatment * t) * log((C0 * (s-1) + exp(gamma_treatment * t) * (lambda^2 - mu + s * (lambda + mu))) / (C0 + exp(gamma_treatment * t) * (lambda + mu))));

% Define the probability generating function for cancerous cells
F_Cancerous = @(s, t) G(s, t);

% Calculation of probabilities using Cauchy integration for cancerous cells
X_Cancerous = 10 * zeros(1601, 1);
P_Cancerous = zeros(1601, 1);

C = [1 + 1i, -1 + 1i, -1 - 1i, 1 - 1i]; % Contour on complex plane
t_val = 10;
for k = 0:1600
    X_Cancerous(k + 1) = k;
    fun_Cancerous = @(s) F_Cancerous(s, t_val) ./ s.^(k + 1);
    P_Cancerous(k + 1) = max(0, real((1 / (2 * pi * 1i)) * integral(fun_Cancerous, (1 + 1i), (1 + 1i), 'Waypoints', C)));
end

% Normalize the PMF
P_Cancerous = P_Cancerous / sum(P_Cancerous);

% Plotting results for cancerous cells
figure
hold on
ax = gca;
ax.FontSize = 14; % Adjust the font size
ax.LineWidth = 2;
plot(X_Cancerous, P_Cancerous, 'color', '#E30B5C', 'linewidth', 2) % Adjust the line width
xlabel('Number of Cancerous Glioma Cells')
ylabel('Probability Mass Function')
title('PMF of Cancerous Glioma Cells')
xlim([0, 50]) % Adjust x-axis limits for better visibility
ylim([0, 0.2]) % Adjust y-axis limits for better visibility
grid on
\end{lstlisting}

\begin{lstlisting}[language=MATLAB, breaklines]
clear all

% Model parameters
lambda = 1/30; % Birth rate
mu = 1/(3.5*24); % Death rate
gamma_treatment = 1/24; % Decay rate of treatment
C0 = 10; % Initial guess for balance term

% Initial values for solving probabilities
n0 = 18;
ntot = 2392.10;
r = 1/(2.5*24);

% Calculation of TA differentiation rate d
syms y s t
d = solve((1 + r/(y - lambda) + r * y / (mu * (y - lambda))) * n0 == ntot, y);
d = double(d);

% Other auxiliary quantities
a = (d - lambda) / mu;

% Define the provided PGF
G = @(s, t) (s.^a * (C0 + lambda + mu)) ./ (-log(C0 * (s-1) + (lambda^2 - mu + s * (lambda + mu)))) .* ...
    (t - exp(gamma_treatment * t) * log((C0 * (s-1) + exp(gamma_treatment * t) * (lambda^2 - mu + s * (lambda + mu))) / (C0 + exp(gamma_treatment * t) * (lambda + mu))));

% Differentiate the PGF with respect to s
dGds = diff(G(s, t), s);

% Evaluate the first derivative at s=1 for different values of t to get the mean
time_values = linspace(0, 100, 1000); % Define the range of time values
mean_values = 10 * double(subs(dGds, {s, t}, {1, time_values}));

% Plotting results for cancerous cells
figure
hold on
ax = gca;
ax.FontSize = 20;
ax.LineWidth = 2;
plot(time_values, mean_values, 'color', '#E30B5C', 'linewidth', 4)
xlabel('Time')
ylabel('Mean number of cells')
title('Cells Over Time')
\end{lstlisting}

\section{MATLAB Code for PDF Graphs}
\label{appendix:MATLAB PDF graphs}

\begin{lstlisting}[language=matlab, breaklines]
    t = 0:0.001:10;

gamma = 0.0028881132525;
C0 = 0.1543;
lambda = 1.025;
mu = 1;
n = 10;

f = (n.*(C0.*exp(-gamma.*t)+lambda+mu)).*exp(-(lambda+mu).*n.*t+(C0.*n.*(exp(-gamma.*t) - 1))./gamma);

semilogx(t, f/100, 'LineWidth', 1.5);
xlabel('Time Delta t');
ylabel('Probability of next event at time Delta t');
title('Log Plot of PDF for the Time of the Next Event: Lipsomal');
legend('Distribution Curve');
\end{lstlisting}

\begin{lstlisting}[language=matlab, breaklines]
    t = 0:0.001:10;

gamma = 0.06931471806;
C0 = 0.1543;
lambda = 1.025;
mu = 1;
n = 10;

f = (n.*(C0.*exp(-gamma.*t)+lambda+mu)).*exp(-(lambda+mu).*n.*t+(C0.*n.*(exp(-gamma.*t) - 1))./gamma);

semilogx(t, f/100, 'LineWidth', 1.5);
xlabel('Time Delta t');
ylabel('Probability of next event at time Delta t');
title('Log Plot of PDF for the Time of the Next Event: Natural');
legend('Distribution Curve');
\end{lstlisting}

\begin{lstlisting}[language=matlab, breaklines]
    %% Natural Cytochalasin Treatment Half Life of 10 minutes

function population_changes = birth_death_simulation_custom(a, t, lambda, mu, C0, gamma)
    % Initialize time and population size
    time = 0;
    population = a;
    population_changes = [time, population];

    % Define the CDF given the population size
    cdf = @(t, population) 1 - exp(-((lambda + mu) * population * t + (C0 * population * (exp(-gamma * t) - 1) / gamma)));

    % Define a range of t values for integration and interpolation
    t_values = 0:0.01:t;

    % Perform simulation until reaching time t
    while time < t
        % Calculate rates of birth and death
        birth_rate = lambda * population;
        death_rate = (mu + (C0*exp(-gamma*time))) * population;

        % Calculate total rate
        total_rate = birth_rate + death_rate;

        % Calculate the CDF for the current population size
        cdf_values = cdf(t_values, population);

        % Ensure the CDF values are strictly increasing and finite
        [cdf_values, unique_idx] = unique(cdf_values); % Remove duplicate values
        t_values_unique = t_values(unique_idx); % Corresponding t values

        % Generate a random number between 0 and 1
        u = rand;

        % Use inverse transform sampling to get delta_t
        delta_t = interp1(cdf_values, t_values_unique, u, 'linear', 'extrap');

        % Update time
        time = time + delta_t;

        % Determine the event (birth or death)
        if rand < birth_rate / total_rate
            population = population + 1; % Birth
        else
            population = population - 1; % Death
        end

        % Record the change in population size
        population_changes = [population_changes; time, population];
        
    end
end

% Parameters
a = 200000; % Initial population size
t = 100; % Simulation time
lambda = 1.025; % Birth rate
mu = 1; % Death rate
gamma = 0.06931471806; % Decay rate of treatment
C0 = .1543; % Initial treatment concentration

num_runs = 10; % Number of simulation runs

% Plotting
figure;
hold on;
count = 1;

for run = 1:num_runs
    % Set random generator seed for each run
    rng(run);
    fprintf("Natural " + count + "\n");
    count = count + 1;
    
    % Run simulation
    simulation_result = birth_death_simulation_custom(a, t, lambda, mu, C0, gamma);
    
    % Extract time and population size from simulation_result
    time = simulation_result(:, 1);
    population = simulation_result(:, 2);
    
    % Plot the population dynamics for this run
    plot(time, population, 'LineWidth', 1.5);
end
hold off;

% Add labels and title
xlabel('Time (Minutes)');
ylabel('Population Size (Cells)');
title('Birth-Death Process Simulation With Natural Cytochalasin D Treatment (10 runs)');
legend('Run 1', 'Run 2', 'Run 3', 'Run 4', 'Run 5', 'Run 6', 'Run 7', 'Run 8', 'Run 9', 'Run 10');
grid on;


%% Liposomal Cytochalasin Treatment Half Life of 240 minutes

% Parameters
a = 200000; % Initial population size
t = 100; % Simulation time
lambda = 1.025; % Birth rate
mu = 1; % Death rate
gamma = 0.0028881132525; % Decay rate of treatment
C0 = .1543; % Initial treatment concentration

num_runs = 10; % Number of simulation runs

% Plotting
figure;
hold on;
count = 1;
for run = 1:num_runs
    % Set random generator seed for each run
    rng(run);
    fprintf("Lipsomal " + count + "\n");
    fprintf("")
    count = count + 1;
    
    % Run simulation
    simulation_result = birth_death_simulation_custom(a, t, lambda, mu, C0, gamma);
    
    % Extract time and population size from simulation_result
    time = simulation_result(:, 1);
    population = simulation_result(:, 2);
    
    % Plot the population dynamics for this run
    plot(time, population, 'LineWidth', 1.5);
end
hold off;

% Add labels and title
xlabel('Time (Minutes)');
ylabel('Population Size (Cells)');
title('Birth-Death Process Simulation With Liposomal Cytochalasin D Treatment (10 runs)');
legend('Run 1', 'Run 2', 'Run 3', 'Run 4', 'Run 5', 'Run 6', 'Run 7', 'Run 8', 'Run 9', 'Run 10');
grid on;
\end{lstlisting}

\section{MATLAB Code for ODE Graph}
\label{appendix:MATLAB ODE graph}

\begin{lstlisting}[language=matlab, breaklines]
    t = 0:0.01:100;

% Parameters
a = 200000; % Initial population size
lambda = 1.025; % Birth rate
mu = 1; % Death rate
gamma1 = 0.06931471806; % Decay rate of treatment (Natural Treatment)
gamma2 = 0.0028881132525; % Decay rate of treatment (Liposomal Treatment)
C0 = 0.1543; % Initial treatment concentration

% Calculate y for the first gamma
y1 = (a./(exp(C0./gamma1))).*exp(t.*(lambda-mu)+((C0.*exp(-gamma1*t))./(gamma1)));

% Calculate y for the second gamma
y2 = (a./(exp(C0./gamma2))).*exp(t.*(lambda-mu)+((C0.*exp(-gamma2*t))./(gamma2)));

% Plotting
plot(t, y1, 'LineWidth', 1.5);
hold on;
plot(t, y2, 'LineWidth', 1.5);
hold off;

xlabel('Time (Minutes)');
ylabel('Population Size (Cells)');
title('ODE Model With Cytochalasin D Treatment');
legend('Natural Treatment', 'Liposomal Treatment');
grid on;

\end{lstlisting}

\end{document}